\documentclass[journal]{IEEEtran}

\usepackage{amsmath}
\usepackage{amssymb}
\usepackage{latexsym}
\usepackage{multirow}
\usepackage{epsfig}
\usepackage{graphics}
\usepackage{mathrsfs}
\usepackage{algorithmic}
\usepackage{algorithm}

\newcommand{\eqdef}{\stackrel{\triangle}{=}}
\newcommand{\beq}{\begin{equation}}
\newcommand{\enq}{\end{equation}}
\newcommand{\ben}{\begin{eqnarray}}
\newcommand{\enn}{\end{eqnarray}}
\newcommand{\bei}{\begin{itemize}}
\newcommand{\eni}{\end{itemize}}

\newcommand{\vsp}{\vspace{.2cm}}

\begin{document}

\title{Optimal Binary/Quaternary Adaptive Signature Design for Code-Division Multiplexing}

\author{Lili~Wei,~\IEEEmembership{Member,~IEEE}, and Wen Chen,~\IEEEmembership{Senior Member,~IEEE}

\thanks{Manuscript received March 13, 2012; revised July 23, 2012 and October 15, 2012; accepted December 11, 2012. The associate editor coordinating the review of this paper and approving it for publication was A. Chockalingam.}
\thanks{The authors are with the Department of Electronic Engineering,
Shanghai Jiao Tong University, China (e-mail: \{liliwei, wenchen\}@sjtu.edu.cn). }
\thanks{This work is supported by the National 973 Project \#2012CB316106 and \#2009CB824904, by NSF China \#60972031 and \#61161130529.}
\thanks{Digital Object Identifier XXXXXXXXXX}}

\markboth{IEEE Transactions on Wireless Communications, VOL.~x,
No.~x, xxxxxx~2013} {Shell \MakeLowercase{\textit{WEI and CHEN}}:
Optimal Binary/Quaternary Adaptive Signature Design for Code-Division Multiplexing}
\maketitle

\begin{abstract}

We consider signature waveform design for synchronous code division multiplexing in the presence of interference and wireless multipath fading channels. The adaptive real/complex signature that maximizes the signal-to-interference-plus-noise ratio (SINR) at the output of
the maximum-SINR filter is the minimum-eigenvalue eigenvector of
the disturbance autocovariance matrix. In digital communication
systems, the signature alphabet is finite and digital signature
optimization is NP-hard. In this paper, first we convert the maximum-SINR objective of adaptive binary signature design into an equivalent minimization problem. Then we present an adaptive binary signature design algorithm based on modified Fincke-Pohst (FP) method that achieves the optimal exhaustive search performance with low complexity. In addition, with the derivation of quaternary-binary equivalence, we extend and propose the optimal adaptive signature design algorithm for quaternary alphabet. Numerical results demonstrate the optimality and complexity reduction of our proposed algorithms.

\end{abstract}

\begin{IEEEkeywords}
Binary sequences, code-division multiplexing, signal-to-interference-plus-noise ratio (SINR), signal waveform
design, signature sets, spread-spectrum
communications.
\end{IEEEkeywords}

\IEEEpeerreviewmaketitle

\section{INTRODUCTION}

\IEEEPARstart{S}{earching} for optimal signature sets has been always with great attention for the growing number of code-division
multiplexing applications such as plain or multiple-input
multiple-output (MIMO) code-division multiple-access (CDMA), multiuser
orthogonal frequency division multiplexing (OFDM), multiuser
ultra-wideband (UWB) systems, etc. In the theoretical context of
complex/real-valued signature sets, the early work of Welch
\cite{1} on total-squared-correlation (TSC) bounds was followed up
by direct minimum-TSC designs \cite{2}-\cite{5} and iterative
distributed optimization algorithms \cite{6}-\cite{8}. Channel and system model
generalizations were considered and handled in \cite{9}-\cite{13}. Signature sets that maximize user capacity are sought in \cite{14}-\cite{15}. Minimum-mean-square-error (MMSE) minimization is used for the design of signature sets for multiuser systems in \cite{16} and over multipath channels in \cite{17}.

All works described above deal with real (or complex) valued signatures, hence their findings constitute only pertinent performance upper bounds for digital communication systems with digital signatures. New bounds on the TSC of {\em binary} signature sets were found \cite{18} that led to minimum-TSC optimal binary signature set designs for almost all signature lengths and set sizes \cite{18}-\cite{20}. The sum capacity, total asymptotic efficiency, and maximum squared correlation of the minimum-TSC binary sets were evaluated in \cite{21}. The sum capacity of other non-minimum-TSC binary sets was calculated in \cite{22} and the user capacity of minimum and non-minimum-TSC binary sets was identified and compared in \cite{23}. The binary code allocation from an orthogonal set is described in \cite{25}. New bounds and optimal designs for minimum TSC quaternary signature sets are derived in \cite{Liming}.

Instead of previous {\em static} binary/quaternary signature design, we consider the NP-hard problem of finding the {\em adaptive} binary/quaternary signature in the code division multiplexing system with interference and multipath fading channels, that maximizes the SINR at the output of the maximum-SINR filter. It is immediately understood that the complex/real minimum-eigenvalue eigenvector of the disturbance autocovariance matrix constitutes an upper bound benchmark. Currently in the literature regarding this problem, direct binary quantization of the minimum-eigenvalue eigenvector is proposed in \cite{Quantized1}-\cite{Quantized2}. The rank-2 proposal that constructs binary signature based on two smallest-eigenvalue eigenvectors is described in \cite{Rank2}. The adaptive binary signature assignment obtained via Euclidean distance minimization from continuous valued arcs of least SINR decrease is presented in \cite{Wei1}.

Different as previous suboptimal approaches, in this paper, first by converting the maximum-SINR objective into an equivalent minimization problem, we propose an adaptive binary signature assignments based on modified Fincke-Pohst (FP) method. The original FP enumeration was proposed in \cite{SD0}, applied to communication system of lattice code decoder in \cite{SD1} as sphere decoding algorithm, and for space-time decoding in \cite{SD2}-\cite{SD3}. Instead of exhaustive searching, FP method considers only a small set of candidate vectors rather than all possible binary points.

In this work, we modify and apply FP method in our adaptive binary signature design to find discrete candidates that lie in a suitable ellipsoid, by a fixed square distance setting with the optimal exhaustive searching results included. Since the radius is fixed for our modified FP algorithm, the complexity uncertainty due to the radius update as shown in the literature of sphere decoding, is not a question in this optimization. In addition, different from communication detection settings, the searching signature candidate set in our algorithm will not expand as signal-to-noise ratio (SNR) increases. We also extend to adaptive signature design with quaternary alphabet, since adaptive quaternary signature design with length $L$ can be proven to be equivalent to adaptive binary signature design with length $2L$. Our contributed binary/quaternary signature design algorithms are guaranteed to find the optimal exhaustive search solutions with much less complexity.

The notations used in this work are as follows. $\{\cdot\}^T$ and $\{\cdot\}^H$ denote the transpose and Hermitian operation respectively.  $\mathbb C^n$ denotes the $n$ dimensional complex field. $Re\{\cdot\}$ and $Im\{\cdot\}$ denote the real part and the imaginary part, and $E\{\cdot\}$ represents statistical expectation. $\mathbf I_n$ denotes the identity matrix of size $n\times n$. We use boldface lowercase letters to denote column vectors and boldface uppercase letters to denote matrices.

The rest of this paper is organized as follows. Section II presents
the system model. The optimal adaptive binary signature assignments based on modified FP algorithm is proposed and described in detail in Section III. Section IV describes the optimal adaptive quaternary signature assignment by quaternary-binary equivalence. Section V is devoted to performance
evaluation. A few concluding remarks are
drawn in Section VI.

%%%%%%%%%%%%%%%%%%%%%%%%%%%%%%%%%%%%%%%%%%%%%%%%%%%%%%%%%%%%%%%%%%%%%%%%%%%%%%%%%%%%%%%%%%

\section{SYSTEM MODEL}

We develop adaptive signature optimization algorithms in the general context of a
synchronous multiuser CDMA-type environment with signature length $L$, where $K$ users transmit simultaneously in frequency and time. Each user transmits over $N$ resolvable multipath fading channels.

Assuming synchronization with the signal of the user of interest $k$,
$k=1,2,\ldots,K$, upon carrier demodulation, chip matched-filtering
and sampling at the chip rate over a presumed multipath extended
data bit period of $L+N-1$ chips, we obtain the received vector ${\bf
r}\in\mathbb{C}^{L+N-1}$ as
\ben
{\bf r} & = & \sqrt{E_k}\;x_{k}{\bf H}_k{\bf s}_k + {\bf z}_k +
{\bf i}_k + {\bf n},
\enn
where $x_k\in\{\pm 1\}$ is the transmitted information bit;
$E_k$ represents transmitted energy per bit period; ${\bf s}_k$ is the signature assigned to user $k$. For binary alphabet ${\bf s}_k\in\{\pm 1\}^L$ while for quaternary alphabet ${\bf s}_k\in \{\pm 1, \pm j\}^L$, with $j\eqdef\sqrt{-1}$.

Channel matrix ${\bf H}_k\in\mathbb{C}^{(L+N-1)\times L}$ for user $k$ is of the form
\beq
{\mathbf H}_{k}\eqdef \left[
\begin{array}{cccc}
h_{k,1} & 0 & \ldots & 0\\
h_{k,2} & h_{k,1} & \ldots & 0\\
\vdots & \vdots &  & \vdots\\
h_{k,N} & h_{k,N-1} &  & 0\\
0 & h_{k,N} &  & h_{k,1}\\
\vdots & \vdots & & \vdots\\
0 & 0 & \ldots & h_{k,N}
\end{array}\right]
\enq
with entries $h_{k,n},n=1,\ldots,N$, considered as complex Gaussian
random variables to model fading phenomena for user $k$ with $N$ resolvable multipaths; ${\bf
z}_k\in\mathbb{C}^{L+N-1}$ represents comprehensively
multiple-access-interference (MAI) to user $k$ by the other $K-1$
users, i.e.
\beq
{\mathbf z}_k\;\eqdef\;\sum_{i=1\;i\ne k}^K
\sqrt{E_i}\;x_{i}{\mathbf H}_i{\mathbf s}_i.
\enq
${\mathbf
i}_k\in\mathbb{C}^{L+N-1}$ denotes multipath induced
inter-symbol-interference (ISI) to user $k$ by its own signal; and
${\bf n}$ is a zero-mean additive Gaussian noise vector with
autocorrelation matrix $\sigma ^2 {\bf I}_{L+N-1}$.

Information bit detection of user \textit{k} is achieved via linear
minimum-mean-square-error (MMSE) filtering (or, equivalently,
max-SINR filtering) as follows
\beq
\hat{x}_{k}= sgn\left(Re\left\{{\bf w}_{MMSE,k}^H {\bf
r}\right\}\right)
\enq
where ${\bf w}_{MMSE,k}\in\mathbb{C}^{L+N-1}$ is
\beq
{\bf w}_{MMSE,k}=c{\bf R}^{-1}{\bf H}_k{\bf
s}_k,
\enq
with $c>0$ and ${\bf R}\eqdef E\{{\bf r} \;{\bf r}^H\}$.

The output SINR of the filter ${\bf w}_{MMSE,k}$ is given by
\ben
SINR_{MMSE,k}({\bf s}_k) & = & \frac{E\left\{\left|{\bf
w}_{MMSE,k}^{H}\left(\sqrt{E_{k}}x_{k}{\bf H}_k{\bf
s}_k\right)\right|^{2}\right\}}{E\left\{\left|{\bf
w}_{MMSE,k}^{H}\left({\bf z}_{k}+{\bf i}_k+{\bf
n}\right)\right|^{2}\right\}}\nonumber\\
 & = & E_{k}{\bf
s}_{k}^{H}{\bf H}_{k}^{H}{\tilde{{\bf R}}}_{k}^{-1}{\bf H}_{k}{\bf
s}_{k}\label{SINR}
\enn
where ${\tilde{{\bf R}}}_{k}\eqdef E\left\{\left({\bf z}_{k}+{\bf i}_k+{\bf
n}\right)\left({\bf z}_{k}+{\bf i}_k+{\bf
n}\right)^H\right\}$ is the autocorrelation matrix of the combined channel disturbance.
For our theoretical developments we disregard the ISI component\footnote{In our simulation studies, the effect of ISI is still taken into account.} and approximate ${\tilde{{\bf R}}}_{k}$ by
\beq
{{\bf
R}}_k\eqdef E\left\{\left({\bf z}_{k}+{\bf n}\right)\left({\bf
z}_{k}+{\bf n}\right)^H\right\}.
\enq

For notational simplicity we define the $L\times L$ matrix
\beq
{\bf Q}_k \eqdef {\bf H}_k^H{\bf R}_{k}^{-1}{\bf H}_{k}.\label{Qk}
\enq
Then, the output SINR in (\ref{SINR}) can be rewritten as
\ben
SINR_{MMSE,k}({\bf s}_k) = E_{k}{\bf
s}_{k}^{H}{\bf Q}_k {\bf
s}_{k}.\label{SINR1}
\enn

Our objective is to find the signature ${\bf s}_k$ that maximizes
$SINR_{MMSE,k}$ of (\ref{SINR1}), in binary alphabet ${\bf s}_k\in \{\pm 1\}^L$ and quaternary alphabet ${\bf s}_k\in \{\pm 1, \pm j\}^L$ respectively.

\bei
\item For binary alphabet ${\bf s}_k\in \{\pm 1\}^L$, let ${\bf Q}_{kR}$ denote the real part of the complex, in
general, hermitian matrix ${\bf Q}_k$, i.e.
\beq
{\bf Q}_{kR}\eqdef Re\{{\bf Q}_k\}.
\enq
The binary signature ${\bf s}_k\in \{\pm 1\}^L$ that maximizes $SINR_{MMSE,k}$ of (\ref{SINR1}) is
equivalent\footnote{Since ${\bf s}\in \left\{\pm 1\right\}^{L}\subset
{\mathbb R}^L$ and ${\bf s}^{H}{\bf Q}_{k}{\bf s}$ is a real scalar,
${\bf s}^{H}{\bf Q}_{k}{\bf s}=Re\left\{{\bf s}^{H}{\bf Q}_{k}{\bf s}\right\}={\bf s}^{T}{\bf Q}_{kR}{\bf s}$} to
\ben
{\bf s}_{k,opt}^{(b)} & = & arg \max_{{\bf s}\in \left\{\pm
1\right\}^{L}}{\bf s}^{T}{\bf Q}_{k}{\bf s}\nonumber\\
& = & arg \max_{{\bf s}\in \left\{\pm
1\right\}^{L}}{\bf s}^{T}{\bf Q}_{kR}{\bf s}.\label{arg_sb}
\enn
The superscript $(b)$ indicates that ${\bf s}_{k,opt}^{(b)}$ is binary.

\item For quaternary alphabet ${\bf s}_k\in \{\pm 1, \pm j\}^L$, the quaternary signature $\mathbf s_k$ that maximizes $SINR_{MMSE,k}$ of (\ref{SINR1}) is given by
\beq
{\bf s}_{k,opt}^{(q)} = arg \max_{{\bf s}\in \left\{\pm
1, \pm j\right\}^{L}}{\bf s}^{H}{\bf Q}_{k}{\bf s}.\label{arg_sq}
\enq
The superscript $(q)$ indicates that ${\bf s}_{k,opt}^{(q)}$ is quaternary.
\eni

A direct approach to these optimization problems (\ref{arg_sb})-(\ref{arg_sq}) will be exhaustive search among all binary/quaternary vectors, specifically $2^L$ candidate vectors for binary optimization and $4^L$ candidate vectors for quaternary optimization. Previous works of \cite{Quantized1}-\cite{Wei1} present some suboptimal approaches. In this work, we will first convert the maximization objective into a minimization problem, and then propose an algorithm based on modified FP method, to searching within a much smaller candidate set with the optimal solution included.

\section{OPTIMAL BINARY SIGNATURE ASSIGNMENT}

\subsection{Formulation}

Regarding the maximum-SINR binary optimization in (\ref{arg_sb}), we first propose to conduct the follow transformation
\ben
{\bf s}_{k,opt}^{(b)} & = & arg \max_{{\bf s}\in \left\{\pm
1\right\}^{L}}{\bf s}^{T}{\bf Q}_{kR}{\bf s}\nonumber\\
& = & arg \min_{{\bf s}\in \left\{\pm
1\right\}^{L}}{\bf s}^{T}\left(\alpha {\bf I}_L - {\bf Q}_{kR}\right){\bf s},\label{arg_sb2}
\enn
where ${\alpha}$ is a parameter greater than the maximum eigenvalue of the matrix ${\bf Q}_{kR}$ and let
\beq
{\bf W} \eqdef \alpha {\bf I}_L - {\bf Q}_{kR}.
\enq
Note that by definition the matrix ${\bf W}$ is Hermitian positive definite.

The Cholesky's factorization of matrix ${\bf W}$ yields ${\bf W} = {\bf B}^T {\bf B}$, where ${\bf B}$ is an upper triangular matrix. Then the binary maximum-SINR optimization in (\ref{arg_sb}) is equivalent to
\ben
{\bf s}_{k,opt}^{(b)} & = & arg \max_{{\bf s}\in \left\{\pm
1\right\}^{L}}{\bf s}^{T}{\bf Q}_{kR}{\bf s}\nonumber\\
& = & arg \min_{{\bf s}\in\{\pm 1\}^L}{\bf s}^T {\bf W} {\bf s}\nonumber\\
& = & arg \min_{{\bf s}\in\{\pm 1\}^L} ||{\bf B}\;{\bf s}||_F^2, \label{arg_sb3}
\enn
where $||\cdot||_F$ denotes the Frobenius norm.

The original Finche-Pohst (FP) method \cite{SD0} searches through the discrete points ${\bf s}$ in the $L$-dimensional Euclidean space which make the corresponding vectors ${\bf z}\eqdef {\bf B} {\bf s}$  inside a sphere of given radius $\sqrt{C}$ centered at the origin point, i.e. $||{\bf B} {\bf s}||_F^2 = ||{\bf z}||_F^2 \le C$. This guarantees that only the points that make the corresponding vectors ${\bf z}$ within the square distance $C$ from the origin point are considered in the metric minimization.

Compared with the original FP method, we have two main modifications: {\em (i)} The original FP algorithm are searching within all integer points, i.e. ${\bf s}\in \mathbb{Z}^L$, while our signature searching alphabet is antipodal binary, i.e. ${\bf s}\in\{\pm 1\}^L$. Hence, the bounds to calculate each entry of the optimal signature are modified, or further tightened, according to our binary searching alphabet to make the algorithm work faster; {\em (ii)} We fix the square distance $C$ setting based on the rank-$1$ approximation ${\bf s}_{\textit{rank-1}}^{(b)}$, which is  the direct sign operator \cite{Quantized1} \cite{Quantized2} on the real maximum-eigenvalue eigenvector of ${\bf Q}_{kR}$, or the rank-$2$ approximation ${\bf s}_{\textit{rank-2}}^{(b)}$ in \cite{Rank2}, or even at higher-rank-optimal solution (rank-3 or rank-4 solution) in \cite{RankD},
\ben
C = \left\{\begin{array}{ll}
{{\bf s}_{\textit{rank-1}}^{(b)}}^T {\bf W}\;{\bf s}_{\textit{rank-1}}^{(b)}  & \quad\quad\textit{if initializing at rank-1}\\
& \quad\quad\textit{approximation}\\
{{\bf s}_{\textit{rank-2}}^{(b)}}^T {\bf W}\;{\bf s}_{\textit{rank-2}}^{(b)}  & \quad\quad\textit{if initializing at rank-2}\\
& \quad\quad\textit{approximation}\\
\quad\quad \cdots\cdots &
\end{array}\right.\label{C}
\enn
such that the searching radius is big enough to have at least one signature point fall inside, while in the meantime small enough to have only a few signature points within. We calculate the ${\bf s}^T {\bf W} {\bf s}$ metric for every signature point ${\bf s}$ that satisfies $||{\bf B} {\bf s}||_F^2 \le C$, such that the optimal signature assignment with minimum ${\bf s}^T {\bf W} {\bf s}$ metric (SINR maximization equivalently) is obtained from the modified FP algorithm directly.

\subsection{Binary Algorithm Derivation}

Let $b_{ij}$, $i,j=1,2,\cdots,L$, denote the entries of the upper triangular matrix ${\bf B}$; let $s_i$, $i=1,2,\cdots,L$ denote the entries of searching vector ${\bf s}$.

According to (\ref{arg_sb3}), the signature points that make the corresponding vectors ${\bf z}={\bf B} {\bf s}$ inside the given radius $\sqrt{C}$ can be expressed as\\
\makebox
{\scalebox{0.85}{%
\parbox{0.5\textwidth}{\ben
{\bf s}^T {\bf W} {\bf s} & = & ||{\bf B}\;{\bf s}||_F^2 = \sum_{i=1}^{L} \left(b_{ii}s_i + \sum_{j=i+1}^L b_{ij}s_j\right)^2\nonumber\\
& = & \sum_{i=1}^{L} g_{ii} \left(s_i + \sum_{j=i+1}^L g_{ij}s_j\right)^2\nonumber\\
& = & \sum_{i=k}^{L} g_{ii} \left(s_i + \sum_{j=i+1}^L g_{ij}s_j\right)^2 + \sum_{i=1}^{k-1} g_{ii} \left(s_i + \sum_{j=i+1}^L g_{ij}s_j\right)^2\nonumber\\
& \le & C\nonumber\\
& & \label{Qs}
\enn}}}\\
where $g_{ii}=b_{ii}^2$ and $g_{ij}=b_{ij}/b_{ii}$ for $i = 1, 2, \cdots,L$, $j = i+1, \cdots, L$.

To satisfy (\ref{Qs}), it is equivalent to consider for every $k = L, L-1, \cdots, 1$,
\beq
\sum_{i=k}^{L} g_{ii} \left(s_i + \sum_{j=i+1}^L g_{ij}s_j\right)^2 \le C. \label{arg_C}
\enq
Then, we can start work backwards to find the bounds for signature entries $s_L, s_{L-1}, \cdots, s_1$ one by one.

We begin to evaluate the last element $s_L$ of the signature vector ${\bf s}$. Referring to (\ref{arg_C}) and let $k=L$, we have
\beq
g_{LL}s_L^2 \le C.\label{sL0}
\enq
Set $\Delta_L=0$, $C_L=C$, we will get
\beq
\left\lceil\; -\sqrt{\frac{C_L}{g_{LL}}} - \Delta_L \; \right\rceil \le s_L \le \left\lfloor\; \sqrt{\frac{C_L}{g_{LL}}} - \Delta_L \;\right\rfloor,\label{sL}
\enq
where $\lceil x \rceil$ is the smallest integer no less than $x$ and $\lfloor x \rfloor$ is the greatest integer no bigger than $x$. As we are searching $s_L\in \{\pm 1\}$, the bounds of $s_L$ in (\ref{sL})
can be modified as
%\beq
%max \left(\left\lceil\; -\sqrt{\frac{C_L}{g_{LL}}} - \Delta_L \; \right\rceil, -1\right) \le s_L \le min \left(\left\lfloor\; \sqrt{\frac{C_L}{g_{LL}}} - \Delta_L \;\right\rfloor, 1\right)\label{sL_modified}
%\enq
\beq
LB_L \le s_L \le UB_L,\label{sL1}
\enq
where
%\ben
%UB_L = min \left(\left\lfloor\; \sqrt{\frac{C_L}{g_{LL}}} - \Delta_L \;\right\rfloor, 1\right),\quad\quad LB_L = max \left(\left\lceil\; -\sqrt{\frac{C_L}{g_{LL}}} - \Delta_L \; \right\rceil, -1\right).\label{sL2}
%\enn
\ben
UB_L & = & \min \left(\left\lfloor\; \sqrt{\frac{C_L}{g_{LL}}} - \Delta_L \;\right\rfloor, 1\right)\nonumber\\
LB_L & = & \max \left(\left\lceil\; -\sqrt{\frac{C_L}{g_{LL}}} - \Delta_L \; \right\rceil, -1\right).\label{sL2}
\enn

For the element $s_{L-1}$ of the signature vector ${\bf s}$, referring to (\ref{arg_C}) and let $k=L-1$, we have
\beq
g_{LL} s_L^2 + g_{L-1,L-1} \left(s_{L-1} + g_{L-1,L}s_L\right)^2 \le C,
\enq
that leads to
\\
\makebox
{\scalebox{0.8}{%
\parbox{0.6\textwidth}{\beq
\left\lceil\; -\sqrt{\frac{C - g_{LL} s_L^2}{g_{L-1,L-1}}} - g_{L-1,L} s_L \;\right\rceil \le s_{L-1} \le \left\lfloor\; \sqrt{\frac{C - g_{LL} s_L^2}{g_{L-1,L-1}}} - g_{L-1,L} s_L \;\right\rfloor.\nonumber
\enq}}}\\
If we denote $\Delta_{L-1} = g_{L-1,L} s_L$, $C_{L-1} = C - g_{LL} s_L^2$ and consider $s_{L-1}\in \{\pm 1\}$, the bounds for $s_{L-1}$ can be expressed as
%\beq
%max \left(\left\lceil\; -\sqrt{\frac{C_{L-1}}{g_{L-1,L-1}}} - \Delta_{L-1} \;\right\rceil, -1\right) \le s_{L-1} \le min \left(\left\lfloor\; \sqrt{\frac{C_{L-1}}{g_{L-1,L-1}}} - \Delta_{L-1} \;\right\rfloor, 1\right). \label{sL-1}
%\enq
\beq
LB_{L-1} \le s_{L-1} \le UB_{L-1},\label{sL-11}
\enq
where\\
\makebox
{\scalebox{0.95}{%
\parbox{0.5\textwidth}{
\ben
UB_{L-1} & = & \min \left(\left\lfloor\; \sqrt{\frac{C_{L-1}}{g_{L-1,L-1}}} - \Delta_{L-1} \;\right\rfloor, 1\right)\nonumber\\
LB_{L-1} & = & \max \left(\left\lceil\; -\sqrt{\frac{C_{L-1}}{g_{L-1,L-1}}} - \Delta_{L-1} \;\right\rceil, -1\right).\quad\quad\quad\label{sL-12}
\enn}}}\\
%\ben
%UB_{L-1} = min \left(\left\lfloor \sqrt{\frac{C_{L-1}}{g_{L-1,L-1}}} - \Delta_{L-1} \right\rfloor, 1\right), LB_{L-1} = max \left(\left\lceil -\sqrt{\frac{C_{L-1}}{g_{L-1,L-1}}} - \Delta_{L-1} \right\rceil, -1\right).\label{sL-12}
%\enn
We can see that given radius $\sqrt{C}$ and the matrix ${\bf W}$, the bounds for $s_{L-1}$ only depends on the previous evaluated $s_L$, and not correlated with $s_{L-2}, s_{L-3}, \cdots, s_1$.

In a similar fashion, we can proceed for $s_{L-2}$ evaluation, and so on.

To evaluate the element $s_k$ of the signature vector ${\bf s}$, referring to (\ref{arg_C}) we will have
\beq
\sum_{i=k}^{L} g_{ii} \left(s_i + \sum_{j=i+1}^L g_{ij}s_j\right)^2 \le C,
\enq
that leads to\\
\makebox
{\scalebox{0.8}{%
\parbox{0.5\textwidth}{\ben
& \left\lceil\; -\sqrt{\frac{1}{g_{kk}} \left(C - \sum_{i=k+1}^L g_{ii}\left(s_i + \sum_{j=i+1}^L g_{ij}s_j\right)^2\right)} - \sum_{j=k+1}^L g_{kj} s_j \;\right\rceil\nonumber\\
& \le s_k \le \left\lfloor\; \sqrt{\frac{1}{g_{kk}} \left(C - \sum_{i=k+1}^L g_{ii}\left(s_i + \sum_{j=i+1}^L g_{ij}s_j\right)^2\right)} - \sum_{j=k+1}^L g_{kj} s_j \;\right\rfloor. \nonumber
\enn}}}\\
If we denote
\ben
\Delta_k & = & \sum_{j=k+1}^L g_{kj}s_j,\nonumber\\
C_k & = & C - \sum_{i=k+1}^L g_{ii}\left(s_i + \sum_{j=i+1}^L g_{ij}s_j\right)^2,
\enn
and take consideration of $s_k\in \{\pm 1\}$, the bounds for $s_k$ can be expressed as
%\beq
%max \left(\left\lceil\; -\sqrt{\frac{C_k}{g_{kk}}} - \Delta_k \;\right\rceil, -1\right) \le s_k \le min \left(\left\lfloor\; \sqrt{\frac{C_k}{g_kk}} - \Delta_k \;\right\rfloor, 1\right). \label{sk}
%\enq
\beq
LB_{k} \le s_{k} \le UB_{k},\label{sk1}
\enq
where
\ben
UB_{k} & = & \min \left(\left\lfloor\; \sqrt{\frac{C_k}{g_{kk}}} - \Delta_k \;\right\rfloor, 1\right),\nonumber\\
LB_{k} & = & \max \left(\left\lceil\; -\sqrt{\frac{C_k}{g_{kk}}} - \Delta_k \;\right\rceil, -1\right).\label{sk2}
\enn
Note that for given radius $\sqrt{C}$ and the matrix ${\bf W}$, the bounds for $s_k$ only depends on the previous evaluated $s_{k+1}, s_{k+2}, \cdots, s_L$.

Finally, we evaluate the element $s_1$ of the signature vector ${\bf s}$. Referring to (\ref{Qs}) and let $k=1$, we will have
\beq
\sum_{i=1}^{L} g_{ii} \left(s_i + \sum_{j=i+1}^L g_{ij}s_j\right)^2 \le C,
\enq
that leads to\\
\makebox
{\scalebox{0.85}{%
\parbox{0.45\textwidth}{\ben
& \left\lceil\; -\sqrt{\frac{1}{g_{11}} \left(C - \sum_{i=2}^L g_{ii}\left(s_i + \sum_{j=i+1}^L g_{ij}s_j\right)^2\right)} - \sum_{j=2}^L g_{1j} s_j \;\right\rceil\nonumber\\
& \le s_1 \le \left\lfloor\; \sqrt{\frac{1}{g_{11}} \left(C - \sum_{i=2}^L g_{ii}\left(s_i + \sum_{j=i+1}^L g_{ij}s_j\right)^2\right)} - \sum_{j=2}^L g_{1j} s_j \;\right\rfloor.\nonumber
\enn}}}\\
If we denote
\ben
\Delta_1 & = & \sum_{j=2}^L g_{1j}s_j,\nonumber\\
C_1 & = & C - \sum_{i=2}^L g_{ii}\left(s_i + \sum_{j=i+1}^L g_{ij}s_j\right)^2,
\enn
and take consideration of $s_1\in \{\pm 1\}$, the bounds for $s_1$ can be expressed as
%\beq
%max \left(\left\lceil\; -\sqrt{\frac{C_1}{g_{11}}} - \Delta_1 \;\right\rceil, -1\right) \le s_1 \le min \left(\left\lfloor\; \sqrt{\frac{C_1}{g_11}} - \Delta_1 \;\right\rfloor, 1\right). \label{s1}
%\enq
\beq
LB_{1} \le s_{1} \le UB_{1},\label{s11}
\enq
where
\ben
UB_{1} & = & \min \left(\left\lfloor\; \sqrt{\frac{C_1}{g_{11}}} - \Delta_1 \;\right\rfloor, 1\right),\nonumber\\
LB_{1} & = & \max \left(\left\lceil\; -\sqrt{\frac{C_1}{g_{11}}} - \Delta_1 \;\right\rceil, -1\right).\label{s12}
\enn

In practice, $C_L$, $C_{L-1}$, $\cdots$, $C_1$ can be updated recursively by the following equations
\ben
\Delta_k & = & \sum_{j=k+1}^L g_{kj}s_j,\\
C_k & = & C - \sum_{i=k+1}^L g_{ii}\left(s_i + \sum_{j=i+1}^L g_{ij}s_j\right)^2\nonumber\\
& = & C_{k+1} - g_{k+1,k+1}\left(\Delta_{k+1} + s_{k+1}\right)^2,
\enn
for $k = L-1, L-2, \cdots, 1$ and $\Delta_L = 0$, $C_L = C$.

The entries $s_L, s_{L-1}, \cdots, s_1$ are chosen as follows: for a chosen candidate of $s_L$ satisfying the bound requirement (\ref{sL1})-(\ref{sL2}), we can choose a candidate of  $s_{L-1}$ satisfying the bounds (\ref{sL-11})-(\ref{sL-12}). If such candidate for $s_{L-1}$ does not exist, we go back and choose other $s_L$. Then search for $s_{L-1}$ that meets the bound requirement (\ref{sL-11})-(\ref{sL-12}) for this new $s_L$. If $s_L$ and $s_{L-1}$ are chosen, we follow the same procedure to choose $s_{L-2}$, and so on. When a set of $s_L, s_{L-1}, \cdots, s_1$ is chosen and satisfies all corresponding bounds requirements, one signature candidate vector ${\bf s}=[s_1, s_2, \cdots, s_L]^T$ is obtained. We record all the candidate signature vectors such that the entries satisfy their bounds requirements and choose the one that gives the smallest ${\bf s}^T {\bf W} {\bf s}$ metric.

Note that this searching procedure will return {\em all} candidates that satisfy ${\bf s}^T {\bf W} {\bf s} \le C$ and gives the one with minimum value. There is at least one vector ${\bf s}_{\textit{rank-D}}^{(b)}$, $D\in\{1,2,3,\cdots\}$ such that its entries satisfy all the bounds requirements, since that is how we set the radius value in (\ref{C}). On the other hand, exhaustive binary search result ${\bf s}_{\textit{exhaustive}}^{(b)}$ will also fall inside the search bounds, since
\beq
{{\bf s}_{\textit{exhaustive}}^{(b)}}^T {\bf W}\; {\bf s}_{\textit{exhaustive}}^{(b)} \le {{\bf s}_{\textit{rank-D}}^{(b)}}^T {\bf W}\;{\bf s}_{\textit{rank-D}}^{(b)} = C.
\enq
Hence, we are guaranteed to find the optimal exhaustive binary search result by the proposed modified FP algorithm with the fixed radius setting of (\ref{C}). Simulation results in Section V also demonstrate this optimality. The setting up choice with different rank initialization will not effect the optimality of the algorithm, but the searching speed will be accelerated with higher rank approximation.

We emphasize that since the radius is fixed for our modified FP algorithm, the complexity uncertainty \cite{SD1}-\cite{complexity1} due to the radius update, which means that the radius need to be expanded if no points found in the sphere and the radius need to be reduced if too many points found within as shown in the literature of sphere decoding, is not a question in this optimization. Also, our proposed searching candidate set will not enlarge as the transmitted energy $E_k$ increase as shown in (\ref{SINR1})-(\ref{arg_sb}).

\subsection{Optimal Binary Algorithm}

We summarize our proposed optimal adaptive binary signature design for (\ref{arg_sb}) in Algorithm 1 as follows.\\

%%%%%%%%%%%%%%%%%%%%%%%%%%%%%%%%%%%%%%%%%%%%%%%%%%%%%%%%%%%%%%%%%%%%%%
%%%%%%%%%%%%%%%%%%%%%%%%%%%%%%%%%%%%%%%%%%%%%%%%%%%%%%%%%%%%%%%%%%%%%%
%%   Binary Signature Design Algorithm II
%%%%%%%%%%%%%%%%%%%%%%%%%%%%%%%%%%%%%%%%%%%%%%%%%%%%%%%%%%%%%%%%%%%%%%
%%%%%%%%%%%%%%%%%%%%%%%%%%%%%%%%%%%%%%%%%%%%%%%%%%%%%%%%%%%%%%%%%%%%%%%
\vspace{-0.2cm}\hspace{-\parindent}\rule{\linewidth}{1pt}\vspace{-0.0cm}
\vspace{-0.0cm}{\bf Algorithm 1}\\
FP Based Binary Signature Design Algorithm\\
\rule{\linewidth}{.5pt}
For the binary signature optimization of ${\bf s}_{k,opt}^{(b)} = arg \max_{{\bf s}\in \left\{\pm
1\right\}^{L}}{\bf s}^{T}{\bf Q}_{kR}{\bf s}$:\vsp\\
\hspace{-\parindent}{\underline{Step 1}}: Let ${\bf q}_{k,1}$ be the real maximum-eigenvalue eigenvector of ${\bf Q}_{kR}$ with eigenvalue ${\lambda_{k,1}}$. Then construct matrix ${\bf W}$ as
\beq
{\bf W} = \alpha {\bf I}_L - {\bf Q}_{kR},\nonumber
\enq
where ${\alpha}$ is a parameter set greater than the maximum eigenvalue of the matrix ${\bf Q}_{kR}$, i.e. $\alpha > \lambda_{k,1}$. Set the square distance based on the rank-D approximation vector ${\bf s}_{\textit{rank-D}}^{(b)}$, $D\in{1,2,3,\cdots}$,
\beq
C = {{\bf s}_{\textit{rank-D}}^{(b)}}^T {\bf W}\; {\bf s}_{\textit{rank-D}}^{(b)}.\nonumber
\enq

\hspace{-\parindent}{\underline{Step 2}}: Operate Cholesky's factorization of matrix ${\bf W}$ yields
\beq
{\bf W} = {\bf B}^T {\bf B},\nonumber
\enq
where ${\bf B}$ is an upper triangular matrix. Let $b_{ij}$, $i,j=1, 2, \cdots,L$ denote the entries of matrix ${\bf B}$. Set
\beq
g_{ii}=b_{ii}^2, \quad\quad g_{ij}=b_{ij}/b_{ii},\nonumber
\enq
for $i = 1, 2, \cdots,L$, $j = i+1, \cdots, L$.

\hspace{-\parindent}{\underline{Step 3}}: Search the candidate vector ${\bf s}$ with entries $s_1, s_2 \cdots, s_L$ according to the following procedure.
\bei
\item[{\em (i)}] Start from $\Delta_L = 0$, $C_L = C$, $metric = C$, ${\bf s}_{min}={\bf s}_{quant}^{(b)}$ and $k = L$.
\item[{\em (ii)}] Set the upper bound $UB_{k}$ and the lower bound $LB_{k}$ as follows
\ben
\left\{\begin{array}{lll}
UB_{k} & = & \min \left(\left\lfloor\; \sqrt{\frac{C_k}{g_{kk}}} - \Delta_k \;\right\rfloor, 1\right),\\
LB_{k} & = & \max \left(\left\lceil\; -\sqrt{\frac{C_k}{g_{kk}}} - \Delta_k \;\right\rceil, -1\right),\\
\end{array}\right.\nonumber
\enn
and $s_k = LB_k -1$.
\item[{\em (iii)}] Set $s_k = s_k + 1$. If $s_k = 0$, set $s_k = 1$. For $s_k \le UB_k$, go to (v); else go to (iv).
\item[{\em (iv)}] If $k=L$, terminate and output $s_{min}$; else set $k = k + 1$ and go to (iii).
\item[{\em (v)}] For $k=1$, go to (vi); else set $k = k - 1$, and
\ben
\left\{\begin{array}{lll}
\Delta_k & = & \sum_{j=k+1}^L g_{kj}s_j, \\
C_k & = & C_{k+1} - g_{k+1,k+1}\left(\Delta_{k+1} + s_{k+1}\right)^2,\\
\end{array}\right.\nonumber
\enn
then go to (ii).
\item[{\em (vi)}] We get a candidate vector ${\bf s}$ that satisfies all the bounds requirements. If ${\bf s}^T {\bf W} {\bf s}\le metric$, then update ${\bf s}_{\min}={\bf s}$ and $metric = {\bf s}^T {\bf W} {\bf s}$. Go to (iii).
\eni

\hspace{-\parindent}{\underline{Step 4}}: Once we get the optimal ${\bf s}_{\min}$ from Step 3 that returns the minimum ${\bf s}^T {\bf W} {\bf s}$ metric, the optimal
adaptive binary signature that maximizes the SINR at the output of MMSE filter is ${\bf s}_{k,opt}^{(b)} = {\bf s}_{\min}$. \\
%\hfill$\blacksquare$
\rule{\linewidth}{.5pt}

\section{OPTIMAL QUATERNARY SIGNATURE ASSIGNMENT}

We extend to consider the adaptive signature design in quaternary alphabet ${\bf s}\in \{\pm 1, \pm j\}^L$ as (\ref{arg_sq}). A heuristic approach will be direct quantization signature vector obtained by applying the sign operator on real part and imaginary part of the complex maximum-eigenvalue eigenvector of ${\bf Q}_k$. However, this is a suboptimal approach and the performance is inferior as shown in simulation section.

In this section, we present a formal procedure of the quaternary-binary equivalence such that the quaternary signature optimization with length $L$ can be equivalent to a binary signature optimization of length $2L$, then the optimal FP Based Binary Signature Design Algorithm proposed in the previous section can be applied directly.

\subsection{Quaternary-Binary Equivalence}

For a quaternary signature ${\bf s}\in \{\pm 1, \pm j\}^L$, we first operate a transform as
\beq
{\bf s} = \frac{1}{2}(1-j) {\bf c},
\enq
such that ${\bf c}\in \{-1-j, -1+j, 1-j, 1+j\}^L$. Note that if the real part and imaginary part of vector ${\bf c}$ are denoted as ${\bf c}_R = Re\{{\bf c}\}$ and ${\bf c}_I = Im\{{\bf c}\}$, this transform will lead to two binary antipodal sequences ${\bf c}_R\in \{\pm 1\}^L$ and ${\bf c}_I\in \{\pm 1\}^L$.

Operate on matrix ${\bf Q}_k$ Cholesky decomposition ${\bf Q}_k = {\bf U}^H {\bf U}$, where ${\bf U}$ is an upper triangular matrix. Then
\ben
{\bf s}^{H}{\bf Q}_{k}{\bf s} & = & \left(\frac{1}{2}(1-j){\bf c}\right)^H{\bf Q}_{k}\left(\frac{1}{2}(1-j){\bf c}\right)\nonumber\\
& = & \frac{1}{2}|| {\bf U} {\bf c} ||_F^2.\label{sq1}
\enn

Define ${\bf y} \eqdef {\bf U} {\bf c}$ and let ${\bf y}_R = Re\{{\bf y}\}$ and ${\bf y}_I = Im\{{\bf y}\}$, ${\bf U}_{R} = Re\{{\bf U}\}$ and ${\bf U}_{I} = Im\{{\bf U}\}$. Then, it is easy to obtain the following equation
\beq
\left[
\begin{array}{c}
{\bf y}_R\\
{\bf y}_I
\end{array}\right] = \left[
\begin{array}{cc}
{\bf U}_{R} & -{\bf U}_{I}\\
{\bf U}_{I} & {\bf U}_{R}
\end{array}\right] \left[
\begin{array}{c}
{\bf c}_R\\
{\bf c}_I
\end{array}\right].\label{y_real}
\enq
Hence, combining equations (\ref{sq1}) and (\ref{y_real}) will lead to\\
\makebox
{\scalebox{0.85}{%
\parbox{0.5\textwidth}{\ben
{\bf s}^{H}{\bf Q}_{k}{\bf s} & = & \frac{1}{2} \left|\left|\left[\begin{array}{cc}{\bf U}_{R} & -{\bf U}_{I}\\{\bf U}_{I} & {\bf U}_{R}\end{array}\right]\left[\begin{array}{c}{\bf c}_R\\{\bf c}_I\end{array}\right] \right|\right|_F^2 \nonumber\\
& = & \underbrace{\left[\begin{array}{c}{\bf c}_R\\{\bf c}_I\end{array}\right]^T}_{\bar{\bf c}^T} \underbrace{ \frac{1}{2} \left[\begin{array}{cc}{\bf U}_{R} & -{\bf U}_{I}\\{\bf U}_{I} & {\bf U}_{R}\end{array}\right]^T \left[\begin{array}{cc}{\bf U}_{R} & -{\bf U}_{I}\\{\bf U}_{I} & {\bf U}_{R}\end{array}\right]}_{\bar{\bf Q}_{kR}} \underbrace{\left[\begin{array}{c}{\bf c}_R\\{\bf c}_I\end{array}\right]}_{\bar{\bf c}},\quad\quad\label{arg_sq1}
\enn}}}\\
where
\beq
\bar{\bf c} \eqdef \left[\begin{array}{c}{\bf c}_R\\{\bf c}_I\end{array}\right]\;\in \{\pm 1\}^{2L},
\enq
is a binary signature with length $2L$. %Note that ${\bf c} = {\bf c}_R + j{\bf c}_I$ is a complex signature with length $L$.

Therefore the quaternary signature optimization with length $L$  in (\ref{arg_sq}) can be transformed into the following binary signature optimization problem with length $2L$
\ben
\bar{\bf c}_{opt}^{(b)} = arg \max_{\bar{\bf c}\in \{\pm 1\}^{2L}} \bar{\bf c}^T \bar{\bf Q}_{kR} \bar{\bf c}.
\enn
After we get the optimal binary sequence $\bar{\bf c}_{opt}^{(b)}$ of length $2L$, split $\bar{\bf c}_{opt}^{(b)}$ into
\beq
\bar{\bf c}_{opt}^{(b)} = \left[\begin{array}{c}{\bf c}_{R,opt}^{(b)}\\{\bf c}_{I,opt}^{(b)}\end{array}\right],
\enq
where ${\bf c}_{R,opt}^{(b)}$ and ${\bf c}_{I,opt}^{(b)}$ are binary sequences in length $L$, i.e. ${\bf c}_{R,opt}^{(b)}\in \{\pm 1\}^L$ and ${\bf c}_{I,opt}^{(b)}\in \{\pm 1\}^L$. Then, the optimal quaternary signature can be constructed as
\beq
{\bf s}_{k,opt}^{(q)} = \frac{1}{2}(1 - j) \left({\bf c}_{R,opt}^{(b)} + j {\bf c}_{I,opt}^{(b)}\right).
\enq

\subsection{Optimal Quaternary Algorithm}

We summarize our proposed optimal adaptive quaternary signature design for (\ref{arg_sq}) in Algorithm 2 as follows.

\vspace{-0.2cm}\hspace{-\parindent}\rule{\linewidth}{1pt}\vspace{-0.0cm}
\vspace{-0.0cm}{\bf Algorithm 2}\\
FP Based Quaternary Signature Design Algorithm\\
\rule{\linewidth}{.5pt}
For the quaternary signature optimization of ${\bf s}_{k,opt}^{(q)}=arg \max_{{\bf s}\in \left\{\pm
1, \pm j\right\}^{L}}{\bf s}^{H}{\bf Q}_{k}{\bf s}$:\vsp

\hspace{-\parindent}{\underline{Step 1}}: We operate on matrix ${\bf Q}_k$ Cholesky decomposition ${\bf Q}_k = {\bf U}^H {\bf U}$. Let ${\bf U}_{R} = Re\{{\bf U}\}$ and ${\bf U}_{I} = Im\{{\bf U}\}$. Construct real matrix $\bar{\bf Q}_{kR}$ as follows
\ben
 \bar{\bf Q}_{kR} = \frac{1}{2} \left[\begin{array}{cc}{\bf U}_{R} & -{\bf U}_{I}\\{\bf U}_{I} & {\bf U}_{R}\end{array}\right]^T \left[\begin{array}{cc}{\bf U}_{R} & -{\bf U}_{I}\\{\bf U}_{I} & {\bf U}_{R}\end{array}\right].\nonumber
\enn

\hspace{-\parindent}{\underline{Step 2}}: Solve the following binary signature optimization problem with signature length $2L$ based on {\em Algorithm 1: FP Based Binary Signature Design Algorithm}
\ben
\bar{\bf c}_{opt}^{(b)} = arg \max_{\bar{\bf c}\in \{\pm 1\}^{2L}} \bar{\bf c}^T \bar{\bf Q}_{kR} \bar{\bf c}.\nonumber
\enn

\hspace{-\parindent}{\underline{Step 3}}: Split
\beq
\bar{\bf c}_{opt}^{(b)} = \left[\begin{array}{c}{\bf c}_{R,opt}^{(b)}\\{\bf c}_{I,opt}^{(b)}\end{array}\right],\nonumber
\enq
where ${\bf c}_{R,opt}^{(b)}$ and ${\bf c}_{I,opt}^{(b)}$ are binary sequences in length $L$. Then, the optimal quaternary signature can be constructed as
\beq
{\bf s}_{opt}^{(q)} = \frac{1}{2}(1 - j) \left({\bf c}_{R,opt}^{(b)} + j {\bf c}_{I,opt}^{(b)}\right).\nonumber
\enq
%\hfill$\blacksquare$
\rule{\linewidth}{.5pt}

As our proposed quaternary signature design algorithm is based on optimal binary signature design algorithm, the optimality can be similarly explained as in previous section.

By using the same quaternary-binary equivalence procedure, we can also extend our previous proposed SDM Based Binary Signature Design Algorithm in \cite{Wei1} to solve the quaternary signature optimization of (\ref{arg_sq}). We denote it as SDM Based Quaternary Signature Design Algorithm with performance comparisons follow in the simulation studies.

We note that the proposed adaptive signature design algorithms for binary and quaternary alphabet can be easily extended to higher-order constellations. For example, for MPSK where each entry of the searching signature $s_k\in [-T,-T+1,\cdots,T-1,T]$, first, the bounds for each entry of the searching signature will not have the $1$, $-1$ constraint; Secondly,
after we get the bound requirement for one entry $s_k$ as $LB_{k} \le s_{k} \le UB_{k}$, the candidate element $s_k$ will be chosen to satisfy this bound requirement and within its alphabet $[-T,-T+1,\cdots,T-1,T]$ over MPSK.

\section{SIMULATION STUDIES}

We first compare performance of adaptive {\em binary} signature assignment algorithms of the following benchmarks: (i) The real maximum-eigenvalue eigenvector  of  ${\bf Q}_{kR} = Re\{{\bf Q}_k\}$, denoted as ''Real max-EV'', which is the theoretical optimal solution over the real field $\mathbb{R}^L$; (ii) The adaptive binary signature assigned by exhaustive search, denoted as ``Exhaustive Binary'', which is the theoretical optimal solution over the binary field $\{\pm 1\}^L$; (iii) The binary signature vector obtained by applying the sign operator on the real maximum-eigenvalue eigenvector of ${\bf Q}_{kR}$, denoted as ``Quantized Binary'' \cite{Quantized1}-\cite{Quantized2}; (iv) The adaptive rank-2 binary signature design algorithm proposed in \cite{Rank2}, denoted as ``Rank2 Binary''; (v) The adaptive binary signature design algorithm in \cite{Wei1} constructing signature vector with slowest descent method (SDM), denoted as ``SDM Based Binary Algorithm''; (v) The optimal adaptive binary signature design algorithm proposed in this work, denoted as ``FP Based Binary Algorithm''. Since different initializing choice from rank-$D$ approximation, $D\in\{1,2,3,\cdots\}$, will not effect the optimality of the algorithm, the simulation curves with those different rank-$D$ setting up actually overlap to one curve. Hence the notation of  ``FP Based Binary Algorithm'' means ``FP Based Binary Algorithm'' with any $C$ setting choice as in (\ref{C}). Same meaning goes to ``FP Based Quaternary Algorithm''.

We consider a code-division multiplexing multipath fading system model with spreading gain $L=16$. Assume that each user's signal
experiences $N=3$ independent fading paths and the corresponding fading channel
coefficients are assumed to be zero-mean complex Gaussian random
variables of equal power, while the additive zero-mean white Gaussian noise is with standard variance. For single user signature assignment performance, the signal power of the user of interest is set to $E_1=10 dB$, while the signal power of present synchronous interferences, $E_2, E_3, \cdots, E_K$ are uniformly spaced between $8 dB$ and $11 dB$. The interfering spreading signatures are randomly generated. For comparison purposes, we evaluate the SINR loss, the difference between SINR of the optimal real signature (Real max-EV) and other adaptive binary signature assignment algorithms. The results that we present are averages over $1000$ randomly generated interferences and channel realizations.

\begin{figure}[!t]
\centerline{\psfig{file=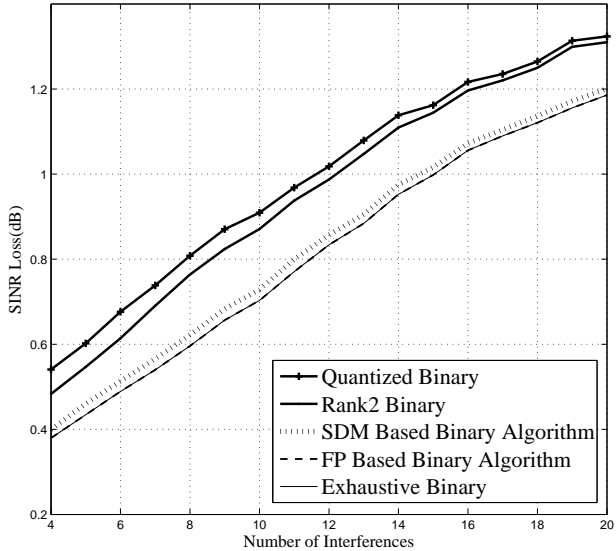,width=3.75in}}
\renewcommand{\baselinestretch}{1}\normalsize
\caption{SINR Loss of various adaptive binary signature assignments versus number of interferences (L=16).}
\end{figure}

\begin{figure}[!h]
\centerline{\psfig{file=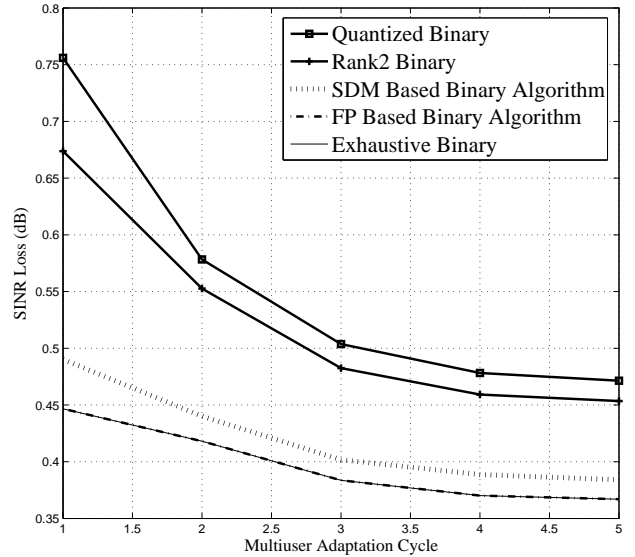,width=3.75in}}
\renewcommand{\baselinestretch}{1}\normalsize
\caption{SINR Loss of various adaptive binary signature assignments versus multiuser adaptation cycle (L=16, K=8).}
\end{figure}

%%%%%%%%%%%%%%%%%%%%%%%%%%
% TABLE 1
%%%%%%%%%%%%%%%%%%%%%%%%%%
\begin{figure*}[!t]
\begin{center}
Table 1: Complexity Comparison: Average Number of Searching Vectors\\
\begin{tabular}{|c|c|c|c|c|c|c|c|c|c|}
\hline
$K$ & 4 & 6 & 8 & 10 & 12 & 14 & 16 & 18 & 20\\
\hline
Proposed with $C_{\textit{rank-1}}$ & 150.21 & 141.87 & 99.91 & 67.32 & 47.47 & 39.24 & 29.84 & 26.15 & 22.36\\
\hline
Proposed with $C_{\textit{rank-2}}$ & 63.87 & 56.35 & 48.41 & 43.30 & 38.53 & 27.53 & 23.71 & 21.07 & 18.24\\
\hline
Proposed with $C_{\textit{rank-3}}$ & 22.11 & 21.75 & 19.71 & 18.23 & 15.64 & 13.20 & 11.90 & 9.69 & 8.59\\
\hline
Exhaustive & 65536 & 65536 & 65536 & 65536 & 65536 & 65536 & 65536 & 65536 & 65536\\
\hline
\end{tabular}\end{center}
\hrulefill
\end{figure*}

In Fig. 1, we plot the SINR loss for binary alphabet as a function of the number of interferences, varying from 4 to 20 interferences. We can observe that SDM based binary algorithm and FP based binary algorithm offer superior performance than the direct quantized binary and rank2 assignments. Furthermore, FP based binary algorithm actually achieves exactly the same optimal exhaustive binary search assignment as we expected.

Then we investigate the multiuser binary signature assignment in a sequential user-after-user manner based on various adaptive binary signature assignments. In such an approach, each user's spreading signature is updated one after the other. Since each spreading signature update results in changes to the interference-plus-noise
statistics seen by the other users, a new update cycle may follow. Several multiuser adaptation cycles are carried out until
numerical convergence is observed. We initialize the signature set arbitrarily and execute one signature set update. In Fig. 2, for a total of $K=8$ users, we plot the SINR loss of one user of interest based on different signature assignment schemes as a function of multiuser adaptation cycle. Still, SDM based binary algorithm and FP based binary algorithm offer superior performance than the direct quantized binary and rank2 assignments. Also, FP based binary algorithm achieves exactly the optimal exhaustive binary search assignment.

\begin{figure}[h]
\centerline{\psfig{file=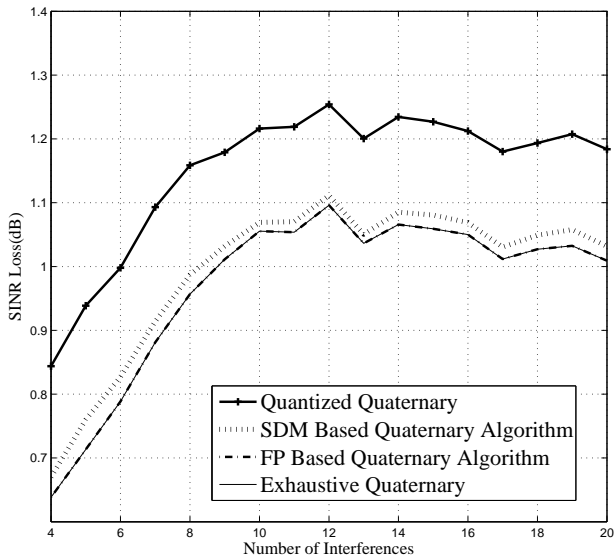,width=3.75in}}
\renewcommand{\baselinestretch}{1}\normalsize
\caption{SINR Loss of various adaptive quaternary signature assignments versus number of interferences (L=8).}
\end{figure}

\begin{figure}[h]
\centerline{\psfig{file=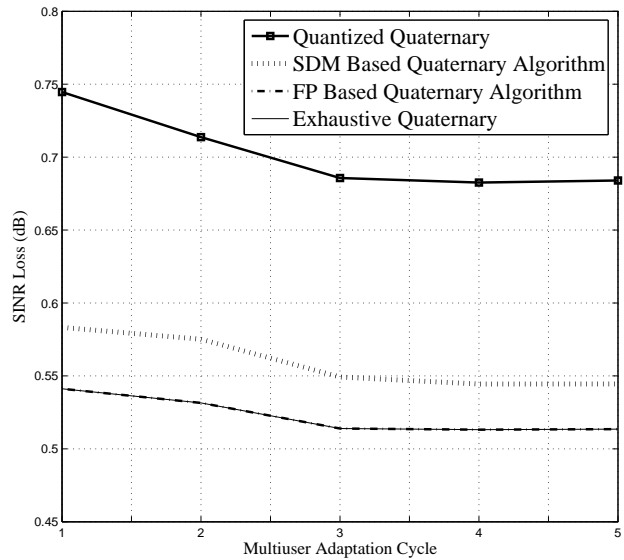,width=3.75in}}
\renewcommand{\baselinestretch}{1}\normalsize
\caption{SINR Loss of various adaptive quaternary signature assignments versus multiuser adaptation cycle (L=8, K=4).}
\end{figure}

We repeat our studies for adaptive {\em quaternary} signature assignment algorithms and compare between the following benchmarks: (i) The complex maximum-eigenvalue eigenvector  of  ${\bf Q}_k$, denoted as ''Complex max-EV'', which is the theoretical optimal solution over the complex field $\mathbb{C}^L$; (ii) The adaptive quaternary signature assigned by exhaustive search, denoted as ``Exhaustive Quaternary'', which is the theoretical solution over the quaternary field $\{\pm 1, \pm j\}^L$; (iii) The quaternary signature vector obtained by applying the sign operator on real part and imaginary part of the complex maximum-eigenvalue eigenvector of ${\bf Q}_{k}$, denoted as ``Quantized Quaternary''; (iv) The adaptive SDM based quaternary signature design algorithm based on the quaternary-binary equivalence procedure and the application of SDM based binary signature assignment in \cite{Wei1}, denoted as ``SDM Based Quaternary Algorithm''; (v) The adaptive quaternary signature design algorithm proposed in this work, denoted as ``FP Based Quaternary Algorithm''. The SINR loss for quaternary assignments are the difference between SINR of the optimal complex signature (Complex max-EV) and other adaptive quaternary assignment algorithms.

We plot the SINR loss for quaternary alphabet as a function of the number of interferences in Fig. 3, and as a function of multiuser adaptation cycle in Fig. 4. We obtain the same results as previous adaptive binary simulations. The SDM based quaternary algorithm and FP based quaternary algorithm offer superior performance than the direct quantized quaternary assignment. Furthermore, our proposed FP based quaternary algorithm actually achieves exactly the optimal exhaustive quaternary search assignment as we expected.

Finally, to demonstrate the complexity reduction of our proposed algorithms with exhaustive search (both return the same optimal results), we compare the statistical average number of binary signature vectors need to be searched to find the optimal solution. For binary exhaustive search, with the setting of $L=16$, the cardinality of the search candidate set will always be $2^L = 65536$. In Table 1 we compare, the statistical average number of binary signature candidate vectors need to be searched. We can see that the candidate set is reduced significantly by our proposed FP based binary signature design algorithm hence lower the complexity dramatically. Also, as we initialize with higher rank approximation, the proposed algorithm is further accelerated.

\section{CONCLUSIONS}

We consider the problem of finding adaptive binary/ quaternary signature in the code division multiplexing system with interference and multipath fading channels, that maximizes the SINR at the output of the maximum-SINR filter. We propose an optimal adaptive binary signature assignments based on modified FP method, that returns the optimal exhaustive searching result with low complexity. In addition, we extend to adaptive quaternary signature assignments and prove that, in general, the adaptive quaternary signature assignment with length $L$ can be equivalent to an adaptive binary signature assignment with length $2L$, hence give the optimal quaternary signature assignments. Simulation studies show the comparisons with our proposed optimal FP based binary/quaternary signature design algorithms, previous suboptimal signature assignments, exhaustive searching and demonstrate the optimality. \\

\end{document}